# Cancer stem cells display extremely large evolvability: alternating plastic and rigid networks as a potential mechanism.
## Network models, novel therapeutic target strategies, and the contributions of hypoxia, inflammation and cellular senescence

Peter Csermely[a,\*], János Hódsági[a], Tamás Korcsmáros[b], Dezső Módos[b,c], Áron R. Perez-Lopez[a], Kristóf Szalay[a], Dániel V. Veres[a], Katalin Lenti[c], Ling-Yun Wu[d], Xiang-Sun Zhang[d]

[a] *Department of Medical Chemistry, Semmelweis University, P.O. Box 260, H-1444 Budapest 8, Hungary*
[b] *Department of Genetics, Eötvös Loránd University, Pázmány P. s. 1C, H-1117 Budapest, Hungary*
[c] *Semmelweis University, Department of Morphology and Physiology, Faculty of Health Sciences, Vas u. 17, H-1088 Budapest, Hungary*
[d] *Institute of Applied Mathematics, Academy of Mathematics and Systems Science, Chinese Academy of Sciences, No. 55, Zhongguancun East Road, Beijing 100190, China*

Cancer is increasingly perceived as a systems-level, network phenomenon. The major trend of malignant transformation can be described as a two-phase process, where an initial increase of network plasticity is followed by a decrease of plasticity at late stages of tumor development. The fluctuating intensity of stress factors, like hypoxia, inflammation and the either cooperative or hostile interactions of tumor inter-cellular networks, all increase the adaptation potential of cancer cells. This may lead to the bypass of cellular senescence, and to the development of cancer stem cells. We propose that the central tenet of cancer stem cell definition lies exactly in the indefinability of cancer stem cells. Actual properties of cancer stem cells depend on the individual "stress-history" of the given tumor. Cancer stem cells are characterized by an extremely large evolvability (i.e. a capacity to generate heritable phenotypic variation), which corresponds well with the defining hallmarks of cancer stem cells: the possession of the capacity to self-renew and to repeatedly re-build the heterogeneous lineages of cancer cells that comprise a tumor in new environments. Cancer stem cells represent a cell population, which is adapted to adapt. We argue that the high evolvability of cancer stem cells is helped by their repeated transitions between plastic (proliferative, symmetrically dividing) and rigid (quiescent, asymmetrically dividing, often more invasive) phenotypes having plastic and rigid networks. Thus, cancer stem cells reverse and replay cancer development multiple times. We describe network models potentially explaining cancer stem cell-like behavior. Finally, we propose novel strategies including combination therapies and multi- target drugs to overcome the Nietzschean dilemma of cancer stem cell targeting: "what does not kill me makes me stronger".

## 1. Cancer as a network development disease

Malignant transformation is increasingly described as a systems-level, network phenomenon. Both healthy and tumor cells can be perceived as networks. Nodes may be the amino acids of cancer-related proteins, where edges are related to secondary chemical bonds. Nodes may also be defined as protein/RNA molecules or DNA-segments, where edges are their physical or signaling contacts. In metabolic networks, nodes are metabolites and edges are the enzymes, which catalyze the reactions to convert them to each other [1–3]. Most of the statements of this review may characterize all these molecular networks of tumor cells and cancer stem cells.

As stated already by Virchow in 1859 [4], cancer is a developmental process. Cancer cells are products of a complex series of cell transformation events. The starting steps are often mutations or DNA-rearrangements, which destabilize the former cellular phenotypes. As a result, a cell population with a large variability in chromatin organization, gene expression patterns and interactome composition is formed [5–9]. In this process, changes in network structure and dynamics play a crucial role.

This review will focus on the large-scale network rearrangements during cancer development—and the emergent, systems-level changes they develop. We will show that changes in network plasticity (and its opposite: network rigidity) may explain central tenets in both cancer development and cancer stem cell behavior. Network plasticity (or in other words network flexibility) can be defined at the level of both network responses and structure [7,10] as it will be detailed in Section 2.

## 2. Malignant transformation proceeds *via* states characterized by increased and decreased network plasticity

### 2.1. Initial increase of network plasticity is followed by a decrease of network plasticity at late stages of carcinogenesis

In our earlier works [3,11,12], summarizing several pieces of evidence we proposed that malignant transformation is a two-phase process, where an initial increase of network plasticity is followed by its decrease at late stages of carcinogenesis. The phenotype of the already established, late-stage cancer cells is still more plastic and immature than that of normal cells, but may often be more rigid than the phenotype of the cells in the intermediate stages of carcinogenesis.

In this concept, network plasticity (or in other words functional network flexibility) can be determined either at level of network responses (network dynamics, attractor structure) and at the level of network structure. The network has a high plasticity at the level of its responses, if small perturbations induce large changes in network structure and dynamics [7,10]. At the level of network, structure network plasticity depends on the internal degrees of freedom of network nodes. Degrees of freedom are reduced by dense clusters, like cliques, or by intra-modular node position. Degrees of freedom are also related to specific network properties: e.g. in transportation-type networks (like metabolic networks) an additional edge may increase the degree of freedom, while in connection type networks (like interactomes) an additional edge may decrease the degree of freedom. The numerical characterization of network plasticity both at the network response and network structure levels is an exciting area of current studies (see more in [10]).

Sources and signs of the initial increase of network plasticity in cancer development are summarized in Table 1 [5,7,13–30]. These sources and signs are related to each other, and might not happen independently. Moreover, Table 1 shows a self-amplified growth of network disorder during tumor development. Self-amplification occurs, when the increased disorder of nodes causes a disorder of their networks, which amplifies the disorder of the nodes further. These synergistic processes cause an accelerated decrease of system-constraints, with a parallel increase in entropy and the degrees of freedom both at the level of the individual nodes and their networks. All these changes lead to the development of more plastic cellular networks in the early phase of cancer. The early stage of cancer development, characterized by an increase of network plasticity, may correspond to the "clonal expansion" phase and the appearance of tumor initiating cells. Such plasticity increase may characterize multiple clonal expansions occurring in some cancer types.

Late stage carcinogenesis is characterized by a decrease of network plasticity reflected by decreasing entropy both at the interactome and signaling network level (such as in case of comparing colon carcinomas to adenomas; Hódsági et al. and Módos et al., unpublished observations). Late stage tumor cells may represent either late stage primary tumor cells, or metastatic cells, which already settled in their novel tissue environment. These findings are in agreement with the recent data of Aihara and co-workers [31,32] showing a transient decrease of entropy of human bio-molecular interaction network during B cell lymphoma, hepatocellular carcinoma and chronic hepatitis B liver cancer development. It is yet to be shown, whether the other types of plasticity increases listed in Table 1 for early stage cancer cells are also reversed in late stage cancer cells.

The dual changes described above correspond well to various steps in the transition to the cancer-specific states, termed as "cancer attractors" by Stuart Kauffman in 1971 [33]. Cancer cells have to first cross a barrier in the quasi-potential (epigenetic) landscape. This barrier might be lowered by mutations or epigenetic changes [5], but its bypass requires a transient destabilization of the transforming cell. This destabilization leads to a more plastic phenotype. This is followed by the stabilization of the cancer cell in the cancer attractor invoking a more rigid phenotype. Importantly, the attractor structure itself may undergo gross changes during cancer development, due to changes in network structure, dynamics and interactions with the environment.

The increase and decrease of network entropy resembles to that observed in cell differentiation processes, where an initial increase of entropy of co-regulated gene expression pattern was followed by a later decrease [34]. An analogous set of events happens in cellular reprogramming, where an early, very heterogeneous, stochastic phase is followed by a late phase, which is programmed by a hierarchical set of transcription factors [35]. Plastic/rigid phenotypes of early/late phases may correspond to the proliferative/remodeling phenotypes of cancer cells obtained by gene expression signature analysis [36]. Importantly, the proliferative/remodeling phenotype duality is very similar to the duality of proliferative/quiescent states of cancer stem cells, which we will discuss in Section 3.

### 2.2. Increase and decrease of network plasticity may alternate in cancer development

The initial increase and later decrease of network plasticity is not displayed uniformly by the heterogeneous cell populations of tumors. Tumors may harbor early phase (plastic) and late phase (rigid) cells at the same time. Importantly, cancer cells in late phase may switch back to an early phase of development [7,26,28]. Thus, tumor cell populations may often be characterized by reversible switches between plastic and rigid network states. Cancer stem cell networks may alternate between plastic and rigid states very intensively, as we will describe in Section 3.

**Table 1**
Sources and signs of increased network plasticity in cancer development.

| Source/sign of increased network plasticity | References | Rationale |
|---|---|---|
| More intrinsically unstructured proteins | [13,16,27] | The increased "conformational noise" of individual proteins makes protein–protein interactions, signaling and metabolism fuzzier in cancer cells |
| Genome instability, chromosomal anomalies | [7] | Destabilization of DNA and chromatin structure, as well as chromosomal anomalies are both consequences and sources of increased system disorder |
| Larger noise of network dynamics (including that of signaling and metabolic networks, as well as larger fluctuations of steady-state values and sensitivity thresholds inducing more stochastic switch-type responses) | [5,14,21] | The larger noise of system dynamics is both a sign of increased network plasticity at the "bottom network" level and a source of further increase in network plasticity at the "top network" level, thus works as a self-amplifier ("bottom networks" describe nodes of the "top networks" like protein structure networks describe nodes of interactomes) |
| Increased entropy of protein–protein interaction and signaling networks | [18–20,23] | The increased entropy of various networks extends the increased disorder of the elementary processes to the system level |
| Larger physical deformability of cancer cells and larger shape heterogeneity | [15,25,30] | Larger cellular deformability shows an increased disorder at the level of the cytoskeletal network, and contributes to chromosomal damage increasing the cellular disorder further. Active mechanisms changing cells from rounded to elongated shape or vice versa increase structural diversity further |
| Alternating symmetric and asymmetric cell division of cancer stem cells | [17,24,29] | Asymmetric cell division increases cellular heterogeneity. Its environmentally regulated alternating character is both a sign and source of increased diversity and plasticity of both intra- and inter-cellular networks |
| Heterogeneous cellular responses to the same stimuli, increased cellular heterogeneity (of tumor cells and infiltrating lymphocytes, macrophages, mast cells, epithelial cells, endothelial cells, fibroblasts and stromal cells) | [7,22,26,28] | Cellular heterogeneity is both a sign of increased plasticity of intra-cellular networks of tumor cells, and, *via* an increase of the heterogeneity of inter-cellular signals of the tumor microenvironment, acts as a source of further increase in system plasticity working as a self-amplifier |

### 2.3. Fluctuating changes of tumor microenvironment may induce an increased adaptation potential of cancer cells via their alternating plastic and rigid network structures

Plasticity and rigidity changes of tumor cell networks are often provoked by changes in the environment of tumor cells. We summarize the major environmental factors enhancing plasticity/rigidity transitions during cancer development in Table 2 [22,26,28,37–64]. As shown in Table 2, inter-cellular networks often develop pro-oncogenic cooperation, but also give a continuously changing environment increasing the adaptation potential of cancer cells. Moreover, tumors grow in a hostile environment characterized by hypoxia, inflammatory responses, low pH, low nutrients, extensive necrosis and targeted by immune and therapeutic attacks. The continuous fluctuation of these stress factors increases the adaptation potential of cancer cells further. Alternating changes in network plasticity and rigidity may play a key role in this "maturation" process. Environment-induced changes may lead to the bypass of cellular senescence and to the development of cancer stem cells (Table 2).

## 3. Network modeling of cancer stem cells

### 3.1. Definition(s) and properties of cancer stem cells

Cancer stem cells have been defined in an American Association for Cancer Research workshop held in 2006 as "cells within a tumor that possess the capacity to self-renew and to cause the heterogeneous lineages of cancer cells that comprise the tumor" [65]. Table 3 lists a number of common beliefs on defining hallmarks of cancer stem cells showing the question marks related to the generalization of these hallmarks [17,26,47,65–78]. We do not consider the uncertainties in the exact definition of cancer stem cells as shortcomings of the field, or as unsolved questions. On the contrary, we think that the very essence of the nature of cancer stem cells is their extreme plasticity, which prevents their precise definition other than 'cells within a tumor that possess the capacity to self-renew and to repeatedly re-build the heterogeneous lineages of cancer cells that comprise a tumor in new environments'. Thus, the central tenet of the definition of cancer stem cells lies exactly in their indefinability. This is why in this review we generally use the term "cancer stem cell" instead of the large variability of names in the literature.

Cancer stem cells have an extremely high adaptation potential to different environments. Thus, the extremely high plasticity of cancer stem cells can be expressed more precisely as an extremely high ability to change the plasticity/rigidity of their networks. Such a property is called metaplasticity in neuroscience [79] and evolvability in genetics [80]. Evolvability (i.e. "*an organism's capacity to generate heritable phenotypic variation*" [80]) is a selectable trait [81], which is mediated by complex networks [82]. This gives us hope that the network requirements of the extremely high evolvability of cancer stem cells can be elucidated and used in anti-cancer therapies in the future. We will summarize the current view on cancer stem cell networks in Section 3.2 and suggest novel therapeutic strategies in Section 4.

Three properties emerge as differentiating hallmarks of cancer stem cells from normal stem cells: (A) increased proliferative potential with a loss of normal terminal differentiation programs; (B) increased individuality (increased niche-independence or niche-parasite behavior); and (C) large efficiency in responses of environmental changes. The high self-renewal potential of normal stem cells gives yet another powerful tool of cancer stem cell survival and evolvability [75,78].

Stemness, especially in tumors, is characterized by large unpredictability, where the outcome is heavily dependent on the individual "stress-history" (past changes of its environment) of the given cancer stem cell. Indeed, more and more data support the early view [83] that cancer stem cells require a number of environmental changes to increase their tumorigenic potential.

- Performing the "defining experiment of cancer stem cells", the serial re-transplantation of tumor cells, for years in *Drosophila* made the cells more tumorigenic with each round of transplantation [84].
- Cancer stem cells could be de-differentiated from differentiated tumor cells in an environment dependent manner [47,71]. Human somatic cells could be reprogrammed to unstable induced epithelial stem cells, which could be transformed to pluripotent cancer stem cells by serial transplantation [76]. Importantly, cancer stem cells were shown to transmit neoplastic properties to normal stem cells [75]. Aged stem cells can be rejuvenated by a systemic young environment [85], which raises the possibility that interactions between quiescent cancer stem cells and their rapidly proliferating environment may help the prolonged survival of cancer stem cells.
- Bioactive food components may foster the aberrant self-renewal of cancer stem cells influencing the balance between their proliferative and quiescent phenotypes [86].
- Anti-cancer therapy often induces a wound-healing response, and contributes to the increase of tumorigenic potential of residual, surviving cancer stem cells [7,28,87].

**Table 2**
Environmental challenges provoking changes in network plasticity and rigidity in cancer development.

| Environmental challenge | Effects on cancer development and on network plasticity/rigidity | References |
| --- | --- | --- |
| Alternating cooperation and competition of inter-cellular networks of the tumor microenvironment | Tumors harbor an extremely heterogeneous cell population of tumor cells, immune cells, epithelial cells, endothelial cells, fibroblasts and stromal cells. These cells may form a cooperating inter-cellular network, and may develop a pro-oncogenic microenvironment. Stabilization of tumor microenvironment shifts cellular networks towards a more rigid state. On the contrary, a fluctuating environment provokes the development of more flexible networks. | [22,26,28,38,49,55,58,64] |
| Changes in the "embeddednes" of tumor cells in the extracellular matrix of tumor microenvironment | Cancer stem cells often occupy a special niche, where niche cells and cancer stem cells mutually help each other's survival by cell adhesion beyond soluble factors. Cancer-associated fibroblasts help the development of a pro-oncogenic microenvironment by a contact dependent mechanism. High molecular weight hyaluronic acid emerges as a key regulator of tumor microenvironment. Increased "embeddedness" of tumor cells in their extracellular environment may increase the rigidity of their networks. (This is the case in the invasive phenotype using extracellular contacts to migrate, and invade new niches, as discussed in Section 3.2). On the contrary, increased independence from the extracellular environment allows increased network plasticity. | [39,42,44,49,52,53,56,61,63] |
| Fluctuations in oxygen tension leading to various degrees of hypoxia and to a large variability of hypoxia-related responses | In agreement with the hypoxia-induced inhibition of senescence, stem cells usually reside in the most hypoxic region of the respective tissue, and are resistant to oxidative stress. Hypoxia-related networks are highly sensitive to minor changes in oxygen concentration, which induces a large variability of hypoxia-related responses. Thus, fluctuations of oxygen tension (which are magnified in tumor microenvironment) may play a key role in the "maturation" of plasticity/rigidity network transitions, and thus the development of cancer stem cells. | [40,41,46,51] |
| Fluctuating intensity of chronic inflammation | Inflammation has a key role in all phases of tumor development. Inflammation may transform stem cells to a more pluripotent, more aggressive phenotype resembling that of cancer stem cells. The Toll-receptor related inflammatory signaling network and its key players, such as MYD88, play an important role in cancer stem cell induction. The fluctuating intensity of chronic inflammations may increase further the adaptation potential of cancer cells mediated by plasticity/rigidity-changes of their networks. | [22,43,57,59,60] |
| Cellular senescence and escape from the senescent state | Although therapy-induced cellular senescence may be beneficial, senescent tumor cells have a number of harmful properties such as: production of inflammatory cytokines and paracrine activators of cancer stem cells, degradation of tumor microenvironment, as well as their potential re-entry to the cell cycle. Escape from the senescent state is a major threat in tumor progression. One of the mediators, survivin, protects senescent cells, may reverse senescence, and characterizes cancer growth showing a co-expression with cancer stem cell markers. Recent studies uncovered a set of highly proliferative microRNAs (such as members of the miR302/367 and miR520 clusters) playing a major role in senescence bypass. Many of these changes characterize cancer stem cells. Cells escaping therapy-induced senescence display a number of proteins characteristic of cancer stem cells such as CD133 or OCT4, and are characterized by an increased amount of antioxidant enzymes. Senescent cells may harbor more rigid networks, while escape from senescence may be characterized by re-gained network plasticity. | [37,45,47,48,50,51,54,62] |

Environmental changes can often be even more potent to increase the tumorigenicity of cancer stem cells, if they have alternating "turn-on" and "turn-off" phases. As an analogy, reversible alternations of stem cell states (with parallel "up" and "down" changes of WNT pathway activity) are required to prevent stem cell senescence [88].
Partly due to the alternating environmental changes some cancer cells reversibly undergo transitions among states that differ in their competence to contribute to tumor growth (e.g. between epithelial and mesenchymal states) [71,74,89]. Proliferative and invasive phenotypes were identified as distinct phenotypes of various cancers [36]. Melanoma cells often switch between these phenotypes [90]. These phenotype transitions, often mediated by genetic lesions, offer an increased chance to adopt a cancer stem cell-like identity [91]. Drug resistance is also a plastic property of some cancer cells. Multi-resistant cancer cell lines reversibly form sensitive or resistant progeny depending on whether the cells are passaged with or without the drug [92]. However, many of the existing evidence for reversible transitions between tumorigenic and non-tumorigenic states comes from studies of cells in culture, and their significance has yet to be established *in vivo* [28].

Alternating and environment-dependent asymmetric and symmetric cell divisions (where template DNA strands are segregated to the daughter cell destined to be a stem cell, or are divided randomly) characterize cancer stem cells. These divisions may be characterized further by the fate of the daughter cells continuing to be stem cells or cells destined to differentiate. Asymmetric lung or gastrointestinal cancer stem cell division is increased by cell density, cell contacts or heat sensitive paracrine signaling, and decreased by hypoxia or serum deprivation [17,24,29]. Asymmetric division characterizes a rigid, quiescent state of cancer stem cells carefully saving encoded information, while separating it from cellular "garbage". On the contrary, symmetric cell division of cancer stem cells characterizes their more plastic, responsive state acquiring and encoding new information. This view is in agreement with the findings that: (A) the tumor suppressor, p53 promoted asymmetric cell division of breast cancer stem cells [93] possibly inhibiting their ability to cycle between asymmetric and symmetric cell division forms; (B) block of asymmetric cell division led to abnormal proliferation and genomic instability in *Drosophila* [94] and (C) rates of asymmetric cell division were different between tumors of a single tumor type [95], and may have to be determined individually for each tumor (or tumor cell).

**Table 3**
Commonly believed cancer stem cell defining hallmarks—and their question marks.

| Commonly believed cancer stem cell defining hallmark | Question marks of hallmark | References |
| --- | --- | --- |
| Is able to form tumors in different environments (especially when using serial transplantation assays) | This definition is the original and most applicable definition of cancer stem cells. Immuno-compromised mice, which are often used in transplantation assay, still may suppress certain tumorigenic cells by a wound-healing type response, and may miss several human cell types (forming the cancer stem cell "niche") needed for tumor development. Mouse-related test environments are not fully informative of tumors in human patients. | [28,65–67,77] |
| Is the precursor of non-tumorigenic cells of the original tumor | This was considered as a consequence of the above definition. However several lines of evidence suggest that the two statements may be unrelated to each other, and the hierarchical lineage typical to most normal stem cells does not characterize many cancer stem cells. Many types of tumors have a surprisingly large genetic heterogeneity suggesting distinct cancer stem cell populations within the same tumor. However, detailed measurement of the tumorigenic potential of the various intra-tumor genotypes is largely missing. Results of the few lineage tracing and selective cell-ablation experiments could not be generalized so far. | [26,28,68,74] |
| Forms only a minor cell subpopulation of tumors | Cancer stem cells may form a rare subpopulation of tumor cells. However, in some experiments a high amount of cancer cells (e.g. 50% instead of the original 0.0004%) were found tumorigenic. | [28,69,70,72–74,77] |
| Is derived from normal stem cells | Though several experiments showed that normal stem cells can be transformed to cancer stem cells, many cancer stem cells seem to be generated from more differentiated cells. | [47,71,75–78] |
| Is resistant to therapy | Therapies inducing cell differentiation successfully target some cancer stem cells. Therapy survival seems to be a stochastic process often depending on *de novo* mutations. | [28] |
| Can be characterized by various cancer stem cell markers | A number of promising cancer stem cell markers (such as CD34, CD38, CD133, CD271, Lgr5) were not generally applicable to the respective tumor type. | [28] |
| Undergoes asymmetric cell division | Cancer stem cells seem to alternate between symmetric and asymmetric cell divisions. This property, which depends on the environment of cancer stem cells, may be an important source of the increased plasticity and evolvability of cancer stem cells. | [17] |

Increase in asymmetric cell division is similar to the development of the quiescent, cooperative state after quorum sensing, while increase in symmetric cell division is similar to quorum quenching [96]. Thus, cancer stem cells may have dedifferentiated to the level that uses community regulation mechanisms similar to those of unicellular organisms, such as bacteria. However, it is an open question, whether inter-cellular cooperation of tumor cells increases with increased asymmetric cell division of cancer stem cells. The recent paper of Dejosez et al. [97] showed the existence of a p53, topoisomerase 1 and olfactory receptor-centered molecular network, which regulates the cooperation of murine induced pluripotent stem cells. It will be a question of later exciting studies to examine whether this molecular network is involved in the maintenance of asymmetric cell division beyond p53, which is a well-known promoter of cell division asymmetry [93].

As a summary, experiments revealed that cancer stem cells are characterized by increased self-renewal potential; increased proliferative potential with a loss of normal terminal differentiation programs; increased individuality and by large efficiency in responses of environmental changes. The appearance and abundance of cancer stem cells seem to be related to the stress-history of the actual tumor. Cancer stem cells have two major phenotypes: a proliferative and a quiescent state characterized by symmetric and asymmetric cell divisions, respectively. More and more studies suggest that the less proliferative cancer stem cell phenotype is characterized by increased invasiveness [98–100]. As we will discuss in Section 3.2, network changes may explain this phenomenon.

In conclusion, as the major hypothesis of this review we propose that the adaptation potential of cancer stem cells is increased by repeated transitions between their plastic (proliferative) and rigid (quiescent/often more invasive) phenotypes. The major driving force behind these transitions is the changes in the microenvironment of cancer stem cells (exemplified by the serial transplantations in the defining experiment of cancer stem cells). Thus, cancer stem cells seem to be able to reverse and replay cancer development multiple times. In cancer development cancer stem cells are repeatedly selected for high evolvability, and became "adapted to adapt" (Fig. 1).

### 3.2. Network models of cancer stem cells

In the last years, several network models of various types of stem cells have been published. We quote here only the work of Muller et al. [101], where gene expression data of approximately 150 of different human stem cell lines were collected and analyzed. Transcriptomes of pluripotent stem cells (embryonic stem cells, stem embryonal carcinomas and induced pluripotent cells) formed a tight cluster, while those of other stem cells were very diverse. Pluripotent stem cells shared a joint sub-interactome, enriched in NANOG, SOX2 and E2F-induced gene-products, as well as in proteins related to "tumorigenesis" in a phenotype analysis [101]. Initial characterization of cancer stem cell-specific networks was performed in osteosarcoma [102]. Gene products of the "stem-transcription factors", NANOG, SOX2, OCT4, KLF4 and MYC, as well as members of the WNT, TGF-β, Notch and Hedgehog pathways emerged as key players of the cancer stem cell signaling network with the concomitant down-regulation of p53, CDKN2A, PTEN, the polycomb repressor complex (including BMI1) and miR-200, where the latter showed an interplay with the ZEB transcriptional repressors. Lacking extensive single cell data on cancer stem cell signaling we currently do not have a clear picture of how many of these mechanisms act simultaneously. However, several lines of evidence suggest that cancer stem cells harbor signaling circuits, which are serially switched on and off in an alternating and cross-regulated manner [76,78,86,103–108].

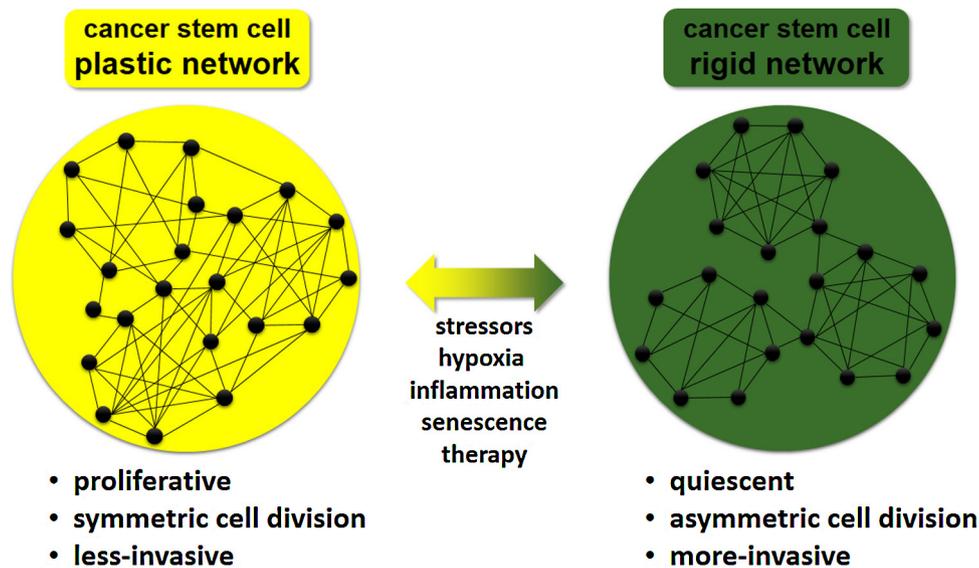

**Fig. 1.** "What does not kill me makes me stronger": Environmental stress-induced alternation of plastic and rigid networks may explain the extremely high evolvability of cancer stem cells. The figure illustrates the major hypothesis of the current review showing that a large variety of environmental stresses (such as hypoxia, inflammation and anti-cancer therapy itself) and stress-responses (such as inflammatory responses and senescence of surrounding cells) may induce the alternation of plastic and rigid states of molecular networks of cancer stem cells. In these networks nodes may be proteins or RNA molecules, while edges may represent their physical or signaling interactions. Similar plasticity/rigidity alterations may occur in metabolic networks (where nodes are small metabolites and edges are enzymes converting them to each other), or in networks of larger cellular complexes, where nodes represent various microfilaments or cellular organelles (such as mitochondria) and edges stand for their interactions. Alternating plastic and rigid network states may help the maturation of cancer stem cells developing their larger adaptability, evolvability and long-term survival. Thus the proverbial saying of Nietzsche: "What does not kill me makes me stronger" [7,125] is especially true for cancer stem cells, and involves the cycling of their molecular networks (such as interactomes, metabolic and signaling networks) between more plastic and more rigid states. Changes of rigid and plastic network structures may involve the creation of new nodes and edges, as well as changes of the weights of existing edges as described in Table 4 in detail.

A number of complex model and real world networks undergo abrupt and extensive changes in their topology as a response to a shortage of resources needed to maintain inter-nodal connections and/or upon environmental stress. These changes are called network topological phase transitions and result in a massive re-organization of network structure, dynamics and function [109–112]. Since the environment of cancers provides an extreme variety of resource/stress levels, which are magnified further by anti-cancer therapies, we have all reasons to assume that networks of cancer cells respond to this environmental variability by topological phase transitions rather often.

Table 4 describes several hypothetical network behaviors, which may explain the extremely high evolvability of cancer stem cells. These mechanisms are interrelated, and may involve the deletion of existing nodes and edges, creation of new nodes and edges, as well as changes of edge weights. Highly dynamic, inter-modular creative nodes may break rigid network structures similarly to lattice defects. Increasing network core size or fuzziness, overlapping, fuzzy network modules or larger intracellular water content may all help this plasticity-increasing process. On the contrary, rigidity-seed nodes may establish a rigid cluster, and rigidity-promoting nodes may help it to grow causing a rigidity phase transition. Decreased dynamics of creative nodes (including chaperones and prions), smaller network core size or more compact network core, separated modules, decreasing intracellular water content may all help this rigidity-increasing process. As a consequence, increase of network plasticity shifts the state-space of the network to a smoother quasi-potential (epigenetic) landscape, while the increase of network rigidity develops a rougher quasi-potential (epigenetic) landscape [3,5,7,10–12,18,82,110,113–124].

These changes are related both to the alternating proliferative and quiescent cancer stem cell phenotypes (Fig. 1) and to the topological network phase transitions mentioned before. Importantly, the quiescent cancer stem cell phenotype often coincides with the invasive phenotype [98–100]. It is rather plausible that high mobility and invasiveness requires a more rigid network structure up to the level of cytoskeletal networks, since physical force required for both migration and invasion can only be exercised efficiently, if the underlying network is not plastic (Fig. 1). However, more studies are needed to assess the generality of the association of the quiescent and the more invasive cancer stem cell phenotypes.

## 4. Network-related drug targeting of cancer stem cells

Cancer stem cells follow Nietzsche's proverbial saying "what does not kill me makes me stronger" [125]. Thus, conventional anti-cancer therapies may actually provoke cancer stem cell development [7,28,87]. However, in the last years a number of cancer stem cell-specific therapies have been developed. Many of the key nodes of cancer stem cell signaling network, such as members of Hedgehog, WNT and Notch pathways, as well as microRNAs, are already used as targets of therapeutic interventions against cancer stem cells. This repertoire is extended by antibodies against putative cancer stem cell markers [78,86]. Differentiation therapy was first suggested by Stuart Kauffman in 1971 [33], and is increasingly used to induce the differentiation of cancer stem cells [78]. Interactions of cancer stem cells with their environment may also provide a panel of targets, like periostin, which is a component of the fibroblast-expressed extracellular matrix [126]. Encouragingly, recent data suggest that some commonly used clinical drugs, such as chloroquine and metformin inhibit the mechanisms used by cancer stem cells to over-ride cellular senescence [50].

Recently a dual drug target strategy, containing the central hit strategy and the network influence strategy, was described to target various diseases [3]. The central hit strategy damages the network integrity of the rapidly proliferating cell in a selective manner. This strategy is useful when attacking plastic networks. The network influence strategy shifts the malfunctioning network of a more differentiated cell back to its normal state. This strategy is useful when attacking rigid networks.

**Table 4**
Hypothetical network-level explanations and possible network modeling approaches of cancer stem cell-like behavior.

| Alternating network property inducing more/less system plasticity | Its possible contribution to cancer stem cell behavior | Possible network modeling approach | References |
|---|---|---|---|
| (A) Changes in network topology and dynamics | | | |
| Successive larger/smaller expression of capacitors of evolvability including creative network nodes | Highly dynamic, inter-modular creative nodes (and/or other capacitor proteins of evolvability including molecular chaperones, prions, and prion-like Q/N-rich proteins) may break rigid network structures similarly to lattice defects. Rigidity-seed nodes may establish a rigid cluster, and rigidity-promoting nodes may help its growth causing a rigidity phase transition. Decreased creative node dynamics may contribute to this process. | Creative nodes can be determined by their inter-modular network position connecting multiple modules at the same time and by their extremely large dynamics. | [11,82,110,113,116,119–122] |
| Pulsation of large/small (or fuzzy/compact) network cores | Larger/fuzzy network cores (i.e. a larger central and dense network segments) increase system robustness, evolvability and plasticity. Decreased network core size and/or increased core compactness may increase the controllability of the system. | Network core size can be calculated by several methods both for undirected and directed (bow-tie) core-periphery networks. | [112] |
| Alternating 'stratus'/'cumulus' (fuzzy/well-separated) network modules | Fuzzy network modules have a large overlap, and form a structure resembling to that of flat, dense, dark, low-lying, stratus clouds. This plastic structure dissipates perturbations well. Well-defined, separated network modules have a structure resembling to that of puffy, white, cumulus clouds. This locally rigid structure develops after stress, and displays reduced perturbation dissipation. | Network modularization methods detecting overlapping modules may be used to assess the separation of network modules. Methods detecting extensive (pervasive) overlaps are especially good for this purpose. | [18,116–118,124] |
| Alternating plastic/rigid network topologies | All the above changes may lead to alternating plastic and rigid network structures. Changing sub-cellular localizations may often contribute to changes in network plasticity/rigidity. | Network rigidity and rigid network clusters may be calculated by generalized pebble game models. | [3,10–12] |
| (B) Changes in network environment | | | |
| Alternating larger/smaller intracellular water content | The more than a dozen known human aquaporin water transporters have differential effects on tumor development. Increased intracellular water may act as "lubricant" increasing system plasticity, while decreased intracellular water content may increase molecular crowding and thus network rigidity. | Intracellular water content may be measured by several experimental methods such as nuclear magnetic resonance spectroscopy. Water-induced changes in molecular network dynamics may be calculated using protein dynamic models and their consequences of interactome, signaling network and metabolic network dynamics. | [114,115] |
| (C) Consequent changes in the quasi-potential (epigenetic) landscape of cancer stem cells | | | |
| Alternating smooth/rough quasi-potential (epigenetic) landscape of cancer cell networks | All the above changes may help the alternation of a smooth quasi-potential (epigenetic) landscape characterized by "shallow" cancer attractors and a rough quasi-potential (epigenetic) landscape characterized by "deep" cancer attractors. | The state space landscape and network attractors can be calculated from analytical models and simulations of network dynamics. | [5,7,118,123,124] |

Since cancer stem cells cycle between plastic and rigid states (Fig. 1, Table 4) conventional chemotherapeutic interventions (targeting plastic networks [3]) only shift the plastic/rigid cycle of cancer stem cells towards the rigid state. For the eradication of cancer stem cells a multi-target therapy is required which attacks their plastic state with "central hit-type" component and their rigid state with a "network influence-type" component at the same time. Successful cancer stem cell eradication requires a network approach all the more, since the network influence strategy requires a multi-target approach by itself [3,123]. Moreover, efficient targeting of rigid networks (corresponding to quiescent cancer stem cells) often requires an indirect approach, where e.g. neighbors of the real target are targeted. These drugs are called allo-network drugs and can also be identified using network-related methods [3,127].

An alternative therapeutic strategy is to "lock" cancer stem cells either in their plastic/proliferative, or in their rigid/quiescent state. Such an intervention has to be followed by a lethal hit of the particular, "locked" state of the cancer stem cell.

We strongly believe that a detailed analysis of cancer stem cell network topology and dynamics based on single cell data will offer a great help to delineate those drug combinations or multi-target drugs, which act minimally on 3 different targets (one attacking the plastic and two the rigid network-related cancer stem cell phenotypes). It will be an important question of further studies, which chronological order and duration of the drug combinations will form the most efficient anti-cancer stem cell therapy.

## 5. Conclusions and perspectives

In conclusion, in this review first we summarized additional pieces of evidence (Table 1) supporting our earlier proposal [3,11,12] that the major trend of malignant transformation is a two-phase process, where an initial increase of network plasticity is followed by a decrease of network plasticity at late stages of carcinogenesis. We showed how environmental changes increase the adaptation potential of cancer cells leading to the bypass of cellular senescence and to the development of cancer stem cells (Table 2). We proposed that cancer stem cells are characterized by an extremely large evolvability helped by their repeated transitions between plastic (proliferative, symmetrically dividing) and rigid (quiescent, asymmetrically dividing, often more invasive) phenotypes (Table 3, Fig. 1). Thus, cancer stem cells reverse and replay cancer development multiple times. Several hypothetical network behaviors were described (Table 4), which may explain the extremely high evolvability of cancer stem cells. Network rigidity/plasticity markers may help to develop novel biomarkers of cancer stem cells. "Locking" cancer stem cells either in their plastic/proliferative, or in their rigid/quiescent state, or combination or multi-target therapies involving a "central hit" and two "network influence" components [3] may overcome the Nietzschean dilemma of cancer stem cell targeting: "what does not kill me makes me stronger".

## Acknowledgements

Work in the authors' laboratory was supported by research grants from the Hungarian National Science Foundation (OTKA-K83314 to P.C.; OTKA-K109349 to T.K.), by the National Natural Science Foundation of China (11131009 and 91330114 to L.W. and X.Z.), and a János Bolyai Scholarship to T.K. P.C. is thankful for the Chinese Academy of Sciences Visiting Professorship for Senior International Scientists supporting the work on this manuscript.

## Appendix A. Supplementary data

Supplementary data associated with this article can be found, in the online version, at doi:10.1016/j.semcancer.2013.12.004.